\documentclass[onecolumn,floatfix,superscriptaddress,a4paper,
               showpacs,showkeys,nofootinbib,preprint]{revtex4}
\usepackage{epsfig}
\usepackage{amssymb,latexsym,amsmath}
\newcommand{\eq}[1]{\begin{align} #1 \end{align}}

\begin{document}

\title{Identity Method for
Particle Number Fluctuations and Correlations}

 \author{M. I. Gorenstein}
 \affiliation{Bogolyubov Institute for Theoretical Physics, Kiev, Ukraine}
 \affiliation{Frankfurt Institute for Advanced Studies, Frankfurt, Germany}

\begin{abstract}
An incomplete particle identification distorts the observed
event-by-event fluctuations of the hadron chemical composition in
nucleus-nucleus collisions.  A new experimental technique called
the {\em identity method} was recently proposed. It eliminated the
misidentification problem for one specific combination of the
second moments in a system of two hadron species. In the present
paper this method is extended to calculate all the second moments
in a system with arbitrary number of hadron species. Special
linear combinations of the second moments are introduced. These
combinations are presented in terms of single-particle variables
and can be found experimentally from the event-by-event averaging.
The mathematical problem is then reduced to solving a system of
linear equations. The effect of incomplete particle identification
is fully eliminated from the final results.

\end{abstract}

\pacs{12.40.-y, 12.40.Ee}

\keywords{hadron chemical fluctuations, incomplete particle
identification, identity method }

\maketitle

\section{Introduction}

A study of event-by-event (e-by-e) fluctuations in high-energy
nucleus-nucleus (A+A) collisions opens new possibilities to
investigate properties of strongly interacting matter (see, e.g.,
review \cite{Koch:2008ia} and references therein). Specific
fluctuations can signal the onset of deconfinement when the
collision energy becomes sufficiently high to create the
quark-gluon plasma~\cite{ood} at the initial stage of A+A
collision.  By measuring the fluctuations, one may also observe
effects caused by dynamical instabilities when the expanding
system goes through the 1-st order transition line between the
quark-gluon plasma and the hadron gas \cite{fluc2}. Furthermore,
the QCD critical point may be signaled by a characteristic
fluctuation pattern \cite{fluc3,Koch:2005vg,Koch:2005pk}.
Fluctuations of the chemical (particle-type) composition of
hadronic final states in A+A collisions are expected to be
sensitive to the phase transition between hadronic and partonic
matter. First data on the e-by-e chemical fluctuations from the
CERN SPS~\cite{Afanasev:2000fu,Alt:2008ca,Kresan:2009qs} and BNL
RHIC~\cite{Abelev:2009if} have been already published, and more
systematic measurements are in progress.
The e-by-e fluctuations of hadron multiplicities have been studied
theoretically in statistical models (see, e.g., Ref.~\cite{CE})
and in dynamical transport models (see, e.g., review \cite{HSD}
and references therein).

Studies of the e-by-e chemical fluctuations assume  particle
number measurements for different hadron species (e.g., pions,
kaons, and protons). The NA49
Collaboration~\cite{Afanasev:2000fu,Alt:2008ca,Kresan:2009qs} has
used the measure $\sigma_{\rm dyn}$, which is defined as the
difference between fluctuations observed in real and mixed events.
The STAR Collaboration~\cite{Abelev:2009if} has used, in addition
to the $\sigma_{\rm dyn}$ measure, the quantity $\nu_{\rm dyn}$
(see, e.g., Ref.~\cite{nudyn}). Moreover, it was suggested long
ago~\cite{Gazdzicki:1997gm,Mrowczynski:1999sf} to quantify
chemical fluctuations by the measure
$\Phi$~\cite{Gazdzicki:1992ri}. Note that different fluctuation
measures can be presented as specific combinations of the second
moments of the multiplicity distribution.
Some important features of different measures for the  e-by-e
fluctuations have been considered in
Ref.~\cite{Gorenstein:2011vq}.

A serious experimental problem of the e-by-e measurements of the
chemical fluctuations is incomplete particle identification; that
is the impossibility to determine uniquely the type of each
detected particle.
The effect of particle misidentifications distorts the measured
fluctuations. For this reason the analysis of chemical
fluctuations is usually performed in a small acceptance, where
particle identification is relatively reliable. However, an
important part of the information on e-by-e fluctuations in  full
phase space is then lost.
Although it is usually impossible to identify each detected
particle, one can nevertheless determine with a high accuracy the
average multiplicities (averaged over many events) for different
hadron species.

In Ref.~\cite{Ga2011} a new experimental technique called the {\it
identity method} was proposed. It solves the misidentification
problem for one specific combination of the second moments in a
system of two hadron species (`kaons' and `pions'). In the present
study we extend these results in two directions. First, we prove
that not only the one specific combination of the second moments
but all the second moments themselves can be uniquely
reconstructed in spite of the effects of incomplete
identification. Second, the identity method is extended to an
arbitrary number $k\geq 2$ of hadron species. This is important
for practical purposes since typically there is an incomplete
identification for pions, kaons, and protons, which means $k=3$.
The identity method is introduced in Section~\ref{sec-id-method}.
In Section~\ref{sec-results} the main results are presented. We
also discuss several examples which illustrate some limiting cases
of particle identification. Section~\ref{sec-sum} presents the
summary.

\section{Identity Method}
\label{sec-id-method}

The identity method was proposed in Ref.~\cite{Ga2011} for the
fluctuation measure $\Psi$.
This method is based on the fact that the analysis of chemical
fluctuations can be performed within two different but fully
equivalent formulations. The first
formulation~\cite{Gazdzicki:1997gm} uses the identity variables;
that is, the $\Psi$ measure of chemical fluctuations is calculated
using single-particle variables $z_i \equiv x_i - \overline{x}$,
where the over-bar denotes averaging over the single-particle
inclusive distribution.
The event variable $Z$, which is a multi-particle analog of $z$,
is defined as $Z \equiv \sum_{i=1}^{N}(x_i - \overline{x})$, where
the sum runs over the $N$ particles in a given event. The measure
$\Psi$ is defined as
\eq{ \label{Psii-def}
\Psi ~\equiv ~ \frac{\langle Z^2 \rangle} {\langle N \rangle}~ -~
\overline{z^2}~,
}
where the symbol $\langle \ldots \rangle$ corresponds to the
e-by-e averaging. One defines the single-particle variable $x_i$
as the {\it identity variable} $w_1(i)$ which equals 1 if the
$i$th particle is of the first type (`kaon'), and $w_1(i) =0$ if
the $i$th particle is  of the second type (`pion'). In a real
measurement, it is unknown exactly whether a given particle is
`kaon' or `pion'. As a consequence of this incomplete
identification the variable $w_1(i)$ is not exactly 0 or 1, but
becomes a distribution function with possible values in the whole
$[0,1]$ interval. Nevertheless, despite the incomplete particle
identification, one can directly use the definition
(\ref{Psii-def}) to evaluate $\Psi$.

In the second formulation, $\Psi$ is calculated in terms of the
moments of the multiplicity distribution. In the case of complete
particle identification it was found~\cite{Mrowczynski:1999sf}
that
\eq{ \label{Psi-moments} \Psi = \frac{1}{\langle N \rangle^3}
\Bigg[\langle N_1^2 \rangle \langle N_2 \rangle^2 + \langle N_1
\rangle^2 \langle N_2^2 \rangle
- 2 \langle N_1 \rangle \langle N_2 \rangle \langle N_1 N_2
\rangle - \langle N_1 \rangle^2 \langle N_2 \rangle - \langle N_1
\rangle \langle N_2 \rangle^2 \bigg]~,
}
where indices 1 and 2 correspond to different hadron species
(`kaon' and `pion'), and $N=N_1+N_2$.
Using the presentation (\ref{Psii-def}), it was shown
\cite{Ga2011} that the measure $\Psi$ can be factorized into a
coefficient that represents the effect of misidentification, and
the quantity (\ref{Psi-moments}), which corresponds to the value
that $\Psi$ would have for complete identification.

We follow Ref.~\cite{Ga2011} and assume that particle
identification is achieved by measuring the particle mass $m$.
Since any measurement is of finite resolution, we deal with
continuous distributions of observed masses  denoted as $\rho_j
(m)$ and normalized as ($j=1,\ldots,k\geq 2$)
\eq{ \label{norm-rho-i} \int dm \,\rho_j (m) = \langle N_j
\rangle~.
}
Note that the functions $\rho_j(m)$  are found for the different
particle species using the values averaged over all particles from
all collision events. The {\it identity variables} $w_j(m)$ will
be defined as
\eq{\label{wi}
w_j(m)~\equiv~\frac{\rho_j(m)}{\rho(m)}~,~~~~~ \rho (m) \equiv
\sum_{i=1}^k\rho_i (m) ~.
}
The {\it complete identification} (CI) of particles corresponds to
distributions $\rho_j (m)$ which do not overlap. In this case,
$w_j = 0$ for all particle species $i\neq j$ and $w_j = 1$ for the
$j$th species. When the distributions $\rho_j (m)$ overlap,
$w_j(m)$ can take the value of any real number from $[0,1]$.

We introduce the quantities $ W_j^2 $, with $j=1,\cdots,k$, and $
W_pW_q$, with $1\le p<q\le k$,
 \eq{\label{Wj2-def}
W_j^2~\equiv~\Big(\sum_{i=1}^{N(n)}w_j(m_i)\Big)^2~,~~~~W_pW_q~\equiv~
\Big(\sum_{i=1}^{N(n)}w_p(m_i)\Big)\times
\Big(\sum_{i=1}^{N(n)}w_q(m_i)\Big)~,
}
and define their event averages  as
\eq{
  \langle W_j^2 \rangle~ =~
\frac{1}{N_{\rm ev}} \sum_{n=1}^{N_{\rm ev}} W_j^2~,~~~~
\langle W_pW_q \rangle ~=~ \frac{1}{N_{\rm ev}} \sum_{n=1}^{N_{\rm
ev}} W_pW_q~,
\label{Wj2-av}
}
where $N_{\rm ev}$ is the number of events, and
$N(n)=N_1(n)+\cdots+N_k(n)$ is the total multiplicity in the $n$th
event. Each experimental event is characterized by a set of
particle masses $\{m_1,m_2,\ldots,m_N\}$, for which one can
calculate the full sets of identity variables:
$\{w_j(m_1),w_j(m_2),\ldots,w_j(m_N)\}$, with $j=1,\ldots,k$.
Thus, the quantities $W_j^2$ and $W_{p}W_q$ are completely defined
for each event, and their average values (\ref{Wj2-av}) can be
found experimentally by straightforward e-by-e averaging. In the
case of CI, one finds  $W_j^2=N_j^2$ and $W_pW_q=N_pN_q$, thus,
Eq.~(\ref{Wj2-av}) yields
\eq{\label{WW}
\langle W_j^2\rangle~=~\langle N_j^2\rangle~,~~~~\langle
W_pW_q\rangle~=~\langle N_pN_q\rangle~.
}

\section{Second Moments of Chemical Fluctuations}
\label{sec-results}

The quantities $\langle W_j^2 \rangle$ and $\langle W_q
W_p\rangle$ can be calculated as follows
\eq{
&\langle W_j^2\rangle ~=~ \sum_{N_1=0}^\infty \sum_{N_2=0}^\infty
\ldots \sum_{N_k=0}^\infty {\cal P}( N_1,\ldots, N_k) \int dm_1^1
P_1 (m_1^1)\ldots  \int dm_{N_1}^1 P_1 (m_{N_1}^1)
 \nonumber\\
 &\times \int dm_1^2 P_2 (m_2^2)\ldots
 \int dm_{N_2}^k P_2(m_{N_2}^2)\times \ldots \times
\int dm_1^k P_k (m_1^k)\ldots \int dm_{N_k}^k P_k (m_{N_k}^k)
\nonumber \\
& \times \Big[w_j(m_1^1) +\cdots  w_j(m_{N_1}^1) + w_j(m_{1}^2) +
\cdots + w_j(m_{N_2}^2)+\ldots + w_j(m_1^k) + \cdots +
w_j(m_{N_k}^k) \Big]^2~ \nonumber \\
& = \sum_{i=1}^k\langle N_i \rangle \big[
u_{ji}^2~-~(u_{ji})^2\big]
%
 + \sum_{i=1}^k\langle N_i^2\rangle (u_{ji})^2
+ 2\sum_{1\leq i<l\leq k}\langle N_i N_l \rangle
u_{ji} u_{jl}~,\label{Wj2}
%
}
\eq{
&\langle W_p W_q\rangle ~=~ \sum_{N_1=0}^\infty
\sum_{N_2=0}^\infty \ldots \sum_{N_k=0}^\infty {\cal P}(N_1,\ldots
,N_k) \int dm_1^1 P_1 (m_1^1)\ldots  \int dm_{N_1}^1 P_1
(m_{N_1}^1)
 \nonumber\\
 &\times \int dm_1^2 P_2 (m_2^2)\ldots
 \int dm_{N_2}^k P_2(m_{N_2}^2)\times \ldots \times
\int dm_1^k P_k (m_1^k)\ldots \int dm_{N_k}^k P_k (m_{N_k}^k)
\nonumber \\
& \times \Big[w_p(m_1^1) +\cdots  w_p(m_{N_1}^1) + w_p(m_{1}^2) +
\cdots + w_p(m_{N_2}^2)+\ldots + w_p(m_1^k) + \cdots +
w_p(m_{N_k}^k) \Big]~ \nonumber \\
& \times \Big[w_q(m_1^1) +\cdots  w_q(m_{N_1}^1) + w_q(m_{1}^2) +
\cdots + w_q(m_{N_2}^2)+\ldots + w_q(m_1^k) + \cdots +
w_q(m_{N_k}^k) \Big]~ \nonumber \\
& = \sum_{i=1}^k\langle N_i \rangle \Big[u_{pqi}
~ -~u_{pi} u_{qi}\Big]
%
 ~+ ~\sum_{i=1}^k\langle N_i^2\rangle u_{pi}
 u_{ki}
 ~ + ~\sum_{1\leq i<l\leq k}\langle N_i N_l \rangle
\Big[u_{pi} u_{ql}~ +~ u_{pl} u_{qi}\Big]~.\label{WpWq}
%
}
In Eqs.~(\ref{Wj2}) and (\ref{WpWq}), ${\cal P}(N_1,\ldots ,N_k)$
is the multiplicity distribution, $P_i (m) \equiv \rho_i(m)/
\langle N_i \rangle $ are the mass probability distributions of
the $i$th species, and ($s=1,2$)
\eq{\label{j-av}
u_{ji}^s ~\equiv~ \frac{1}{\langle N_i \rangle } \int dm \,
w_j^s(m)~ \rho_i (m)  ~,~~~~ u_{pqi}~\equiv~\frac{1}{\langle
N_i\rangle} \int dm\, w_p(m)\, w_q(m)~\rho_i(m)~.
}

In the case of CI,  when the distributions $\rho_j (m)$ do not
overlap, one finds that
\eq{\label{CI-j}
u_{ji}^{s} ~ =~\delta_{ji}~,~~~~~ u_{pqi}~ =~ 0~,
}
and Eqs.~(\ref{Wj2}) and (\ref{WpWq}) reduce then to
Eq.~(\ref{WW}). The incomplete particle identification transforms
the second moments $\langle N_j^2\rangle$ and $\langle
N_pN_q\rangle$ to the quantities $\langle W_j^2\rangle$ and
$\langle W_p W_q\rangle$, respectively. Each of the later
quantities contains  linear combinations of all the first and
second moments, $\langle N_i\rangle$ and $\langle N_i^2\rangle$,
as well as all the correlation terms $\langle N_i N_l \rangle$.
Having introduced the notations
\eq{
\langle W_j^2 \rangle  - \sum_{i=1}^k\langle N_i \rangle \big[
u_{ji}^2~-~ (u_{ji})^2\big] ~\equiv~b_j~,
%
~~~~ \langle W_pW_q\rangle -\sum_{i=1}^k\langle N_i \rangle \big[
u_{pqi}~ -~ u_{pi} u_{qi} \big] ~\equiv~b_{pq}~, \label{bpq}
}
one can transform Eqs.~(\ref{Wj2}) and (\ref{WpWq}) to the
following form:
\eq{
&\sum_{i=1}^k\langle N_i^2\rangle~ u_{ji}^2
+ 2\sum_{1\leq i<l\leq k}\langle N_i N_l \rangle ~ u_{ji} u_{jl}
~=~b_j~,~~~~j=1,2,\ldots,k~,\label{bj1}\\
& \sum_{i=1}^k\langle N_i^2\rangle ~u_{pi} u_{qi}
+ \sum_{1\leq i<l\leq k}\langle N_i N_l \rangle
\Big(u_{pi}u_{ql}~+~ u_{pl} u_{qi}\Big)~=~b_{pq}~,~~~~ 1\leq
p<q\leq k~.\label{bpq1}
}
The right-hand side of Eqs.~(\ref{bj1}) and (\ref{bpq1}) defined
by Eq.~(\ref{bpq}) are experimentally measurable quantities. The
same is true for the coefficients  $u_{ji}^{s}$ (with $s=1$ and
$2$) entering the left-hand side of Eqs.~(\ref{bj1}) and
(\ref{bpq1}).
Therefore, Eqs.~(\ref{bj1}) and (\ref{bpq1}) represent a system of
$k+ k(k-1)/2$ linear equations for the $k$ second moments $\langle
N_j^2\rangle$ with $j=1,\ldots,k$ and $k(k-1)/2$ correlators
$\langle N_pN_q\rangle$ with $1\leq p< q\leq k$.

In order to solve Eqs.~(\ref{bj1}) and (\ref{bpq1}) we introduce
the $[k+k(k-1)/2]\times[k+k(k-1)/2]$ matrix $A$
\eq{\label{A}
A~=~
\begin{pmatrix}
a_{1}^{1} & \ldots & a_{1}^{k} &|&
a_{1}^{12}
&\ldots  & a_{1}^{(k-1)k}\\
. & . & . &|&
 . &  .  & .\\
. & . & . &|&
 . &  .  & .\\
 a_{k}^{1} & \ldots & a_{k}^{k} & | &
 a_{k}^{12}
 &\ldots &  a_k^{(k-1)k}  \\
--- & --- &--- &|&---& ---&---
\\
a_{12}^1 &\ldots  & a_{12}^k  & | & a_{12}^{12}
& \ldots & a_{12}^{(k-1)k}\\
. & . & . &|&
. & . & .  \\
. & . & . &|&
 . & . & .  \\
a_{12}^{k}& \ldots & a_{(k-1)k}^k  & | &   a_{(k-1)k}^{12} &
\ldots & a_{(k-1)k}^{(k-1)k}
 \end{pmatrix}~,
}
where
\eq{
a^{i}_{j}~&\equiv ~ u_{ji}^2~,~~1\leq i,j\leq k~;~~~~~ a_{i}^{pq}
\equiv ~2 u_{ip}u_{iq}~,~~ 1\leq p< q\leq k~,~~
i=1,\ldots ,k ~~;\\
a_{pq}^{i}~&\equiv ~u_{pi}u_{qi}~,~~1\leq p< q\leq k~,~~i=1,\ldots
,k~;
\\
 a_{pq}^{lm}~&\equiv~u_{pl}u_{qm}+u_{ql}u_{pm}~,~~
1\leq p<q\leq k~,~~1\leq l<m\leq k~.
}
The solution of Eqs.~(\ref{bj1}) and (\ref{bpq1}) can be presented
by Cramer's formulas in terms of the determinants
\eq{\label{Nj2-sol}
\langle N_j^2\rangle ~=~ \frac{{\rm det}~A_j}{{\rm
det}~A}~,~~~~~\langle N_pN_q\rangle ~=~ \frac{{\rm
det}~A_{pq}}{{\rm det}~A}~,~
}
where the matrices $A_j$ and $A_{pq}$ are obtained by substituting
in the matrix $A$ the column $ a_{1}^{j},\ldots, a_{k}^{j},
a_{12}^j,\ldots, a_{(k-1)k}^j$ and the column $a_1^{pq},\ldots ,
a_k^{pq}, a^{pq}_{12},\ldots , a_{(k-1)k}^{pq}$, respectively, for
the column $b_1,\ldots, b_k,$ $b_{12}, \ldots, b_{(k-1)k}$.
Therefore, if ${\rm det}A\neq 0$, the system of linear equations
(\ref{bj1}) and (\ref{bpq1}) has a unique solution (\ref{Nj2-sol})
for all the second moments.
In the case of CI (\ref{CI-j}), one finds ${\rm det}A=1$, ${\rm
det}A_j=b_j$,  and ${\rm det}A_{pq}=b_{pq}$. The solution
(\ref{Nj2-sol}) reduces then to Eq.~(\ref{WW}).

Introducing the  $[k+k(k-1)/2]$-vectors
\eq{\label{vectors}
{\cal N}~\equiv~
\begin{pmatrix}
\langle N_1^2\rangle  \\
\ldots \\
\langle N_k^2\rangle \\
\langle N_1N_2\rangle \\
\ldots\\
\langle N_{k-1}N_k \rangle
\end{pmatrix}
~,~~~~~~
 {\cal B}~\equiv~
\begin{pmatrix}
b_1  \\
\ldots \\
b_k \\
b_{12} \\
\ldots\\
b_{(k-1)k}
\end{pmatrix}~,
}
one can write Eqs.~(\ref{bj1}) and (\ref{bpq1}) in the matrix form
$A{\cal N} ={\cal B}$. The solution (\ref{Nj2-sol}) can be then
rewritten as
\eq{\label{AN}
{\cal N}~=~A^{-1}~{\cal B}~,
}
where $A^{-1}$ is the inverse matrix of $A$.
For two particle species, $k=2$, this solution takes the form
\eq{\label{A2}
\begin{pmatrix}
\langle N_1^2 \rangle \\
\langle N_2^2 \rangle \\
\langle N_1 N_2 \rangle
\end{pmatrix}
~=~
\begin{pmatrix} u_{11}^2 ~~~~& u_{12}^2~~~~& 2 u_{11} u_{12} \\
 u_{21}^2~~~~& u_{22}^2~~~~
&2 u_{21} u_{22} \\
u_{11} u_{21} ~~~~~ & u_{12} u_{22}~~~~ & u_{11} u_{22} + u_{12}
u_{21}
\end{pmatrix} ^{-1}~
\begin{pmatrix}
b_1\\
b_2\\
b_{12}
\end{pmatrix}
~.
}
Then Eq.~(\ref{A2}) yields
\eq{
 \langle N_1^2\rangle
~&=~ \frac{b_1u_{22}^2~+~b_2u_{12}^2~-~2b_{12}
u_{12}u_{22}}{\big(u_{11}u_{22}~-~u_{12}u_{21}\big)^2}~,
\label{N1-sol}
\\
 \langle N_2^2\rangle ~& = ~
 \frac{b_2u_{11}^2~+~
 b_1u_{21}^2 ~-~ 2b_{12}u_{21}u_{11}}
 {\big( u_{11}u_{22}~-~u_{12}u_{21}\big)^2}~,\label{N2-sol}\\
\langle N_1N_2\rangle ~& = ~
\frac{b_{12}\big(u_{11}u_{22}+u_{12}u_{21}\big) -b_1u_{22}u_{21}-
b_2u_{11}u_{12}}{\big(u_{11}u_{22}~-~u_{12}u_{21}\big)^2}~.
\label{N12-sol}
}
These results, inserted into Eq.~(\ref{Psi-moments}), provide an
alternative way to evaluate $\Psi$ directly from the moments of
the multiplicity distribution.

In general, the particle-by-particle identification is difficult;
that is, it is not known whether a given particle really
corresponds to the $j$th sort. On the other hand, the statistical
identification in terms of the functions $\rho_j(m)$ is usually
reliable. Experimental measurements of the $\rho_j(m)$ functions
give the average numbers of each particle species. In most cases,
a unique calculation of the second moments using
Eq.~(\ref{Nj2-sol}) is also possible.
There is,  however, an extreme situation when the only available
experimental information consists of the average particle
multiplicities. This leads to {\em random identification} (RI)
which only defines, for each particle, the probabilities $p_j$ of
being of the $j$th sort. These probabilities are evidently equal
to $p_j=\langle N_j \rangle/\langle N\rangle$, where
$N=\sum_{i=1}^k N_i $. This situation is described by the mass
distributions given by
\eq{ \label{ri}
\rho_j(m)~ = ~ \langle N_j \rangle~f(m)~,
}
where $\int dm f(m)=1$; that is, all functions $\rho_j(m)$ have
the same shape $f(m)$ but different normalization $\langle
N_j\rangle$. With these distributions one finds
%
\eq{
u_{ji}~=~
\frac{\langle N_j\rangle }{\langle N\rangle}
~\equiv~p_j~,~~~~j=1,\ldots,k~.
}
This leads to ${\rm det}A=0$, and Eqs.~(\ref{bj1}) and
(\ref{bpq1}) do not define the second moments in a unique way. In
fact, from Eq.~(\ref{Wj2-def}) follows
\eq{\label{Wri}
\langle W_j^2\rangle ~=~p_j^2~\langle N^2 \rangle ~,~~~~
\langle W_{pq}\rangle ~=~p_pp_q ~\langle N^2\rangle~,
}
that is, in the case of  RI, the measured values (\ref{Wri})
include only the average multiplicities $\langle N_j\rangle$ and
the second moment of the total multiplicity $\langle N^2\rangle$.
Equations (\ref{bj1}) and (\ref{bpq1}) for $\langle N_j^2\rangle,$
and $\langle N_pN_q\rangle$ reduce to a single relation
\eq{\label{Nri}
\sum_{j=1}^k\langle N_j^2\rangle~ +~2 \sum_{1\leq p<q\leq
k}\langle N_pN_q\rangle ~=~\langle N^2\rangle~,
}
where the right-hand side of Eq.~(\ref{Nri}) is the only
experimentally measured combination of the second moments.
Therefore, RI gives only one restriction on $k+ k(k-1)/2$ second
moments and thus admits an infinite number of solutions for
$\langle N_j^2\rangle$ and $\langle N_pN_q\rangle $. Any correctly
normalized multiplicity distribution ${\cal P}(N_i,\ldots,N_k)$,
which reproduces experimental values of the first moments, would
reproduce Eqs.~(\ref{bj1}) and (\ref{bpq1}); that is, in the case
of RI the experimental data do not provide any non-trivial
information on chemical fluctuations.

It is also instructive to consider an illustrative example when
particle species are divided into two groups: $j=1,\ldots,k_R$
with  RI (\ref{ri}) and $j=k_R+1,\ldots,k$ with CI (\ref{CI-j}).
Equations (\ref{bj1}) and (\ref{bpq1}) are then given by
\eq{\label{R-RI}
 \langle N_R^2\rangle~&=~\langle W_1^2\rangle =\ldots =\langle
W_{k_R}^2\rangle =\langle W_{1}W_{2}\rangle= \ldots =\langle
W_{k_R-1}W_{k_R}\rangle  ,~~~~N_R\equiv \sum_{j=1}^{k_R}~N_j~;\\
\langle N_R N_q \rangle ~&=~\langle
W_{1}W_{q}\rangle=\ldots=\langle
W_{k_R}W_{q}\rangle~,~~~~k_R+1\leq q~\leq k~; \label{Rq-RI}
}
\eq{\label{j2-CI}
 \langle N_j^2\rangle~&=~\langle
W_j^2\rangle~,~~~~j=k_R+1,\ldots,k~;\\
\langle N_{p}N_q\rangle~&=~\langle W_{p}W_{q}\rangle~ ,~~~~
j=k_R+1,\ldots,k~;~~~~ k_R+1\leq p<q\leq k~ . \label{pq-CI}
}
Quantities $\langle W_j^2\rangle$ and $\langle W_pW_q\rangle$ can
be measured experimentally by using their definitions according to
Eq.~(\ref{Wj2-def}). For the particle species with RI, as follows
from Eqs.~(\ref{R-RI}) and (\ref{Rq-RI}), the second moments of
particle multiplicities include only their total multiplicity
$N_R$. Therefore, one knows all individual average multiplicities
$\langle N_j\rangle$, but as far as chemical fluctuations are
concerned, all particles in the RI group, $1\leq j\leq k_R$, look
undistinguishable. On the other hand, this fact does not prevent
calculations of the second moments (\ref{j2-CI}) and (\ref{pq-CI})
in the  CI group, $k_R+1\leq j \leq k$.

In the formulation considered in this paper, the functions
$\rho_j(m)$ are defined as  quantities averaged over all particles
from all collision events. One can consider the set of events with
fixed total multiplicity $N$.  All formulae of this paper
straightforwardly apply in this case too. The only modifications
are: 1) event averaging $\langle \ldots \rangle$ over all events
is changed to averaging $\langle \ldots \rangle_N$ over events
with fixed $N$; 2) the functions $\rho_j(m)$ should be replaced by
$\rho_j(m;N)$ calculated for fixed $N$. This procedure may open
some new possibilities in the studies of chemical fluctuations.

\section{Summary}
\label{sec-sum}
An incomplete particle identification prevents a straightforward
measurement of the second moments $\langle N_j^2\rangle$ and
$\langle N_pN_q\rangle$ of the multiplicity distribution. In this
paper we extend the identity method proposed in
Ref.~\cite{Ga2011}. We introduce the quantities $\langle
W_j^2\rangle$ and $\langle W_pW_q\rangle$. Each of these
quantities is a specific linear combination of all the first and
second moments $\langle N_i\rangle$ and $\langle N_i^2\rangle$, as
well as the correlation terms $\langle N_iN_l\rangle$. The
quantities $\langle W_j^2\rangle$ and $\langle W_pW_q\rangle$ are
presented in terms of e-by-e averages of functions depending on
the single-particle identity variables according to
Eq.~(\ref{Wj2-av}), and can thus be measured experimentally.
Mathematically, the problem  of finding all the second moments
$\langle N_j^2\rangle$ and $\langle N_pN_q\rangle$ is then reduced
to solving the system of $k+k(k-1)/2$ linear equations (\ref{bj1})
and (\ref{bpq1}). All coefficients entering the left-hand side of
these equations are given in terms of experimentally measurable
density functions $\rho_j(m)$. The right-hand side in
Eqs.~(\ref{bj1}) and (\ref{bpq1}) is defined by Eq.~(\ref{bpq})
which also includes experimentally measurable quantities.
In most cases the determinant of matrix (\ref{A}) is not equal to
zero and, therefore, all second moments of particle number
distributions can be uniquely reconstructed by Eq.~(\ref{Nj2-sol})
from event-by-event measurements despite the effects of incomplete
identification. This is valid for an arbitrary number $k\geq 2$ of
different hadron species.  The matrix $A^{-1}$ in Eq.~(\ref{AN})
represents the correction of the measured values (\ref{bpq}). Such
a correction eliminates the effect of misidentification. This
provides the values of all the second moments $\langle
N_j^2\rangle$ and $\langle N_pN_q\rangle$ in a model-independent
way, as they would be obtained in an experiment in which each
particle is uniquely identified.
However, all measured quantities entering Eqs.~(\ref{bj1}) and
(\ref{bpq1}) contain experimental errors.  Therefore, the
practical applicability of the identity method procedure
constructed in this paper requires further studies.

\begin{acknowledgments} I am thankful to V.V.~Begun, E.L.~Bratkovskaya,
M.~Ga\'zdzicki, W.~Greiner, M.~Hauer, M. Ma\' ckowiak,
St.~Mr\'owczy\'nski, A.~Rustamov, and P.~Seyboth for fruitful
comments and discussions. This work was in part supported by the
Program of Fundamental Research of the Department of Physics and
Astronomy of NAS, Ukraine.
\end{acknowledgments}

\end{document}